\documentclass[aps,prb,twocolumn,superscriptaddressgroupaddress,floatfix]{revtex4-2}
\usepackage{amssymb}
\usepackage{physics}
\usepackage{graphicx}
\usepackage{times}
\usepackage{amsmath}
\usepackage{amsthm}
\usepackage{amsfonts}
\usepackage[T1]{fontenc}
\usepackage[latin9]{inputenc}
\usepackage{array}
\usepackage{multirow}
\usepackage{color}
\usepackage{bm}
\usepackage{color}

\usepackage{bbm}
\usepackage{hyperref}
\usepackage{babel}
\usepackage{float}
\usepackage{hyperref}
\hypersetup
{	colorlinks,%
	citecolor=green,%
	linkcolor=red,%
	urlcolor=blue%
}
\allowdisplaybreaks[4]
\begin{document}
\title{Avalanche stability transition in interacting quasiperiodic systems}
        \author{Yi-Ting Tu}
	\author{DinhDuy Vu}
	\author{S. Das Sarma}
	\affiliation{Condensed Matter Theory Center and Joint Quantum Institute, Department of Physics, University of Maryland, College Park, Maryland 20742, USA}
	\begin{abstract}
Coupling a 1D quasiperiodic interacting system to a Markovian bath, we study the avalanche instability of the many body localized phase numerically, finding that many body localization (MBL)
is more stable in pseudorandom quasiperiodic systems than the corresponding randomly disordered systems for a disorder strength $W>8$, potentially up to arbitrarily large system sizes.
  We support our conclusion by additionally developing real space RG arguments, and provide a detailed comparison between quasiperiodic and random MBL from the avalanche instability perspective, concluding that the two belong to different universality classes.
	\end{abstract}

	\maketitle

	\section{Introduction}
        Many-body localization (MBL)~\cite{Abanin2019,Nandkishore2015} is a manifestation of the generic failure of the eigenstate thermalization hypothesis (ETH)~\cite{Deutsch1991,Srednicki1994} and as such violates the basic premise of quantum statistical mechanics in that an isolated generic interacting quantum system does not thermalize.  This is a lofty claim, and the subject has remained controversial since the evidence for MBL has been mostly through small system numerical simulations, and an intricate mathematical proof for arbitrarily large disorder~\cite{Imbrie2016}. Does MBL really exist as a thermodynamic phenomenon or is it just like a glass with very slow relaxation with no eventual violation of ergodicity?

        An important conceptual advance was recently made  in the subject by Morningstar et al.~\cite{Morningstar2022}, see also \cite{Sels2021}, where the issue was inverted asking the question which mechanism(s) could conceivably destabilize MBL (and how), assuming its existence at very large disorder in the thermodynamic limit.  The issue can be studied by coupling the system weakly to a Markovian bath, and then carefully following the system dynamics to discern whether the bath induced thermal inclusions grow ("avalanche") or stabilize with system size and disorder.  The conclusion of Refs.~\cite{Morningstar2022,Sels2021} for interacting randomly disordered 1D problem is that thermodynamic MBL is definitely destabilized for disorder strength $W<18$, by contrast, all existing studies of MBL find a critical disorder $\sim5$ for the MBL transition, showing that the putative MBL so far studied in the literature is only a finite-size MBL or at best a crossover phenomenon.
In the existing MBL simulations, the relaxation is slow, but the system would thermalize eventually with increasing system size, and at best, a finite-size-MBL is being observed.

        Since the studies of Refs.~\cite{Morningstar2022,Sels2021} are restricted only to the interacting Anderson model (i.e.\ random disorder), a natural question arises about the nature of MBL in the interacting incommensurate quasiperiodic system (e,g.\ Aubry-Andr\'e model~\cite{Aubry1980}) where the disorder is strictly deterministic (and not random at all).  This is important conceptually since the nature of single-particle localization is qualitatively different in 1D random and quasiperiodic models (e.g. the random model is always localized whereas the quasiperiodic model allows for extended or localized single-particle states depending on the disorder strength), and also from an experimental perspective, since all substantive experimental studies of 1D MBL have focused on the quasiperiodic systems (and not on random systems) making the understanding of MBL in quasiperiodic systems of considerable empirical significance~\cite{Luschen2018,Kohlert2019,Michael2015,Rispoli2019,Alexander2019,An2018}. 

        In the current work, we do not assume that the thermodynamic MBL is entirely driven by avalanche instability. Instead, we consider avalanche as a possible thermodynamic correction to the finite-size MBL. The logic can be stated explicitly in the random model. For relatively small sizes, MBL transition takes place at $W\sim 3.5$. As the system size grows, large thermal seeds become statistically important and may destroy MBL. This means that the thermodynamic MBL transition happens at a much stronger disorder than the finite-size counterpart in the random model. Can we make the same statement about quasiperiodic models? Specifically, can an avalanche happen, and if it happens, is it as effective as one in the random model? These two questions cannot be addressed definitely by either analytical or numerical tools (in the thermodynamic sense), so we construct our argument based on two approaches and try to draw a qualitative picture. 
        	
        In the first approach, we study the thermalization by coupling the spin chain to an ideal thermal bath, which can be regarded as an infinitely large thermal seed. Within this setting, we find the avalanche is less effective in the quasiperiodic model, yielding a lower critical potential strength than the random case. This represents the worst-case avalanche, and does not reflect the realistic situation because a real thermal seed is always finite and gets less efficient as the thermal region grows. To estimate how large a thermal seed can be, we rely on the second approach by coarse-graining the lattice and using the block decimation scheme. This is where we see the drastic difference between the two models. In the former, an arbitrarily large (proportional to the system size) thermal seed can always exist, inducing an avalanche similar to the worst-case scenario (induced by an ideal thermal bath). On the other hand, for the quasiperiodic model, our RG shows that the size of thermal seeds is capped and not diverging with the system size, meaning that the realistic avalanche is far less severe than the worst estimation. Combing these two findings, we can predict that the finite-size MBL in quasiperiodic models is less likely to undergo a drastic destabilization as seen in the random case. As such the MBL transition in thermodynamic limit might not be much higher than that for finite-size systems. In short, in the thermodynamic limit, interactions make the Anderson disordered system much more ergodic than the corresponding quasiperiodic system where the MBL appears to be much more stable (although the single-particle localization is less stable since the Aubry-Andr\'e model allows for extended states at weak disorder). In terms of utility,  our work also suggests that large-size experiments that require MBL may benefit from quasi-periodic potentials.
        
         This paper is organized as follows: In Sec.~\ref{sec:theory}, we introduce the Hamiltonians we study, and provide the level-statistics data for finite-size MBL. In Sections~\ref{sec:open} and \ref{sec:rg}, we discuss and present the results of the open system simulation and the real space renormalization approach for thermodynamic MBL, respectively. Finally, we present our conclusion in Sec.~\ref{sec:conclusion}. Additional details and results are available in the appendices.

        \section{Theory}\label{sec:theory}
	We consider the Heisenberg model with onsite potential, which is an open 1D spin-$\frac{1}{2}$ chain with the Hamiltonian
	\begin{equation}
	\label{eq:quasiperiodic_hamiltonian}
	H = \frac{1}{4} \sum_{j=1}^{L-1} \vec\sigma_j \cdot \vec\sigma_{j+1} + \frac{1}{2} \sum_{j=1}^L h_j Z_j,  
	\end{equation}
	where $\vec{\sigma}_j = (X_j, Y_j, Z_j)$ are the Pauli matrices on site $j$. We consider two models for the onsite potential: (i) quasi-periodic potential $h_j=W\cos(2\pi\varphi j +\phi)$ with the golden ratio $\varphi=(1+\sqrt{5})/2$, and (ii) random potential where $h_j$ is a random variable drawn from a uniform distribution $[-W,W]$. For the former, we average over an ensemble of random $\phi$; while for the latter, the average is performed over random configurations. The quasi-periodic potential case is our new result compared to Refs.~\cite{Morningstar2022,Sels2021}, which only study the random model. In the following, we demonstrate and then explain that the difference between these two models is not simply quantitative but is actually qualitative, and reflects distinguishable physics. 
	
\begin{figure}
    	\centering
    	\includegraphics[width=\columnwidth]{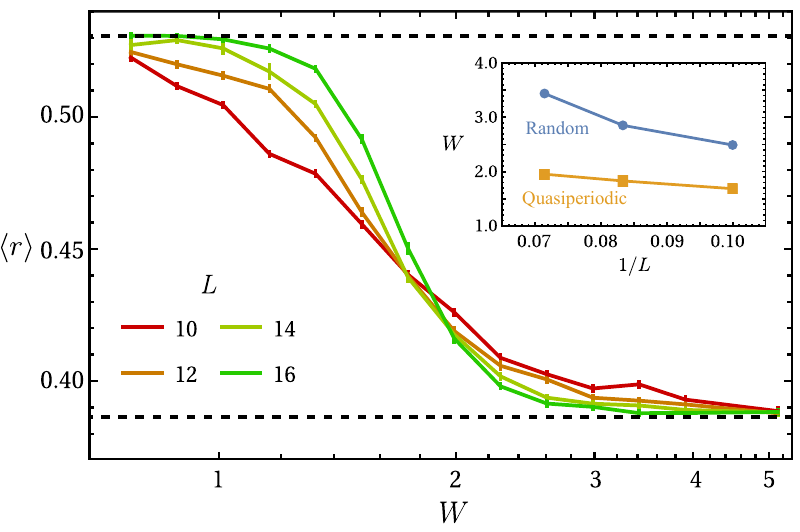}
        \caption{The mean level spacing ratio for the quasi-periodic model versus disorder strength. The grid lines denote the GOE limit $\langle r \rangle=0.5307$ and the Poisson limit $\langle r \rangle=0.3863$. The inset shows the intersection of $\langle r \rangle$ in $W$ between $L$ and $L+2$ drifting to larger value as the system size increases. The data for the random model used to calculate the intersections are from Ref.~\cite{Pal2010}.} \label{levelstatistics}. 
    \end{figure}

    Before we discuss the two main approaches for thermodynamic MBL,
    we provide the level statistics data as a benchmark for finite-size MBL transition in the quasi-periodic spin chain in Fig.~\ref{levelstatistics}. The mean level spacing ratio is computed by averaging $r_n=\text{min}(\delta_n,\delta_{n-1})/\text{max}(\delta_n,\delta_{n-1})$ over the ordered spectrum with $\delta_n=E_n-E_{n-1}$ being the spacing between two neighboring energy levels. We use periodic boundary condition, and $r$ is averaged over the entire spectrum of the zero-magnetization sector, as well as many different initial phases. As we increases the disorder strength, $\langle r \rangle$ changes from $0.5307$ for the GOE (ETH) to $0.3863$ for the Poisson distribution (MBL) \cite{Atas2013}. In the range $L=10-16$, the finite-size MBL transition happens at $W\sim 1.7-2.0$ for the quasi-periodic model and $\sim 2.5-3.5$ for the random model.
    
        \section{Open system decay rate}\label{sec:open}
        In this section, we identify the avalanche landmark for the quasiperiodic model, following
        the approach in \cite{Morningstar2022,Sels2021}. We also calculate that for the random model (with part of the data taken from \cite{Morningstar2022}) as a comparison. We couple the first spin to an ideal thermal bath, so that the spin chain of small size $L$ can be used to simulate the induced thermalization behavior near a thermal inclusion in a large MBL system. The dynamics of this small spin chain is described by the Lindblad equation
	
	\begin{equation}
	\label{eq:lindbladian}
	\mathcal{L}[\rho] = -i[H,\rho] + \gamma\sum_{\mu}\left(L_{\mu}\rho L^\dagger_\mu - \frac{1}{2}\{L^\dagger_\mu L_\mu,\rho\}\right),
	\end{equation}
        where $\gamma>0$ is the bath coupling strength
        and $L_\mu=(X_1,Y_1,Z_1)$.
        We then denote $\lambda_1$ as the eigenvalue of $\mathcal{L}$ with the second largest real part as zero is always the eigenvalue with the largest real part, corresponding to the maximally mixed steady state. $\tau=-1/\mathrm{Re}(\lambda_1)$ is the life time of the slowest decay mode and can be considered as the lower bound of the time scale in which the (non-ideal) thermal bath in the large MBL system thermalizes a spin at distance $L$ away. 

In the context of avalanche instability, this small chain thermalization describes the expansion of a thermal bubble in an MBL system.
  Suppose there is a thermal bubble of length $l$ in an MBL system and our length-$L$ chain simulates its neighbor on one side. If the bubble thermalizes the chain, the thermal region grows to the size $l+2L$, including the other side.
  We can estimate its level spacing to be $1/2^{l+2L}$, whose reciprocal is the longest time scale at which the bubble can be treated as an effective thermal reservoir.
  Therefore, if the decay time $\tau$ grows faster than $4^L$, this thermal region stops growing at some finite $L$, and since thermal seeds are rare to begin with in the MBL phase, the system remains localized.
Otherwise, the bubble can expand to fill the entire system, causing thermalization and destroying the finite-size MBL.
Therefore, we display our decay rate in the scale of $4^{-L}$. This is the same strategy as in Refs.~\cite{Sels2021,Morningstar2022}, now adapted for a quasiperiodic ``disorder''.

The coupling to the ideal thermal bath provides a worst-case estimation to study the stability, and we will show that, even in this situation, the quasiperiodic spin chain is more stable.
  Whether the thermal seeds occur naturally in the quasiperiodic spin chain (as in the random spin chain) is another question, which we discuss in Sec.~\ref{sec:rgdetails}.
  Besides simulating the rare thermal region, deliberate planting of thermal inclusions leading to quantum avalanches has recently been observed in interacting quasiperiodic 1D atomic systems~\cite{Leonard2020}, providing a direct motivation for our open system calculation.

\subsection{Calculation method}
\begin{figure}[h]
	\centering
	\includegraphics[width=\columnwidth]{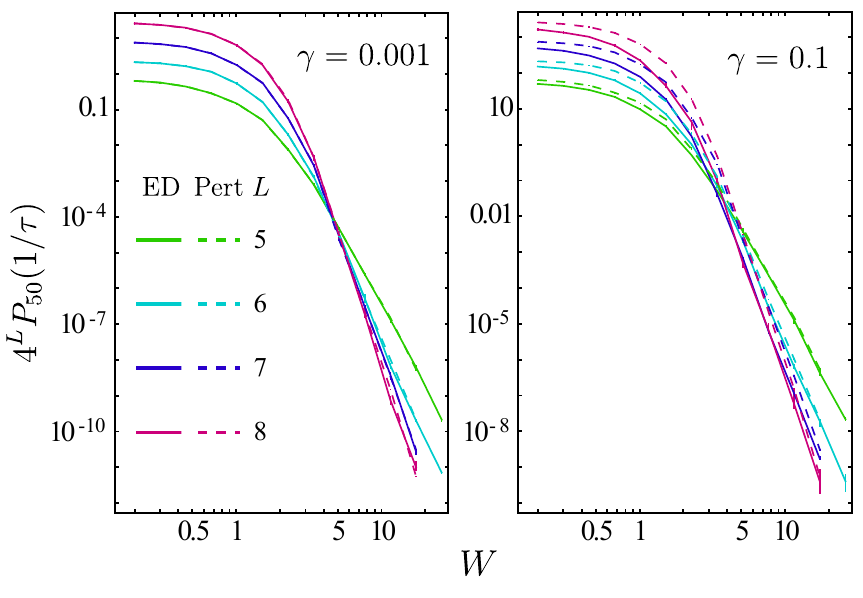}
	\caption{Scaled decay rate computed from exact diagonalization (solid lines) and perturbation theory (dash lines) in the quasi-periodic model. \label{fig1}}
	\end{figure}

          Since the exact diagonalization (ED) of $\mathcal{L}$ (a $4^L\times4^L$ sparse matrix) is numerically heavy, we first discuss whether the perturbation method (as used in Ref.~\cite{Sels2021,Morningstar2022}) can be used in our case to calculate the decay rate.
        We compare the ED result (of the quasiperiodic spin chain with small $L$) with the first-order perturbation ($\gamma\ll 1$) approach, in which $\lambda_1$ equals the second largest eigenvalue of the projected Lindbladian
	\begin{equation}
	\mathcal{L}_{nm}=\langle m|\mathcal{L}[|n\rangle\langle n|]|m\rangle\,,
	\end{equation}
	a $2^L\times 2^L$ (non-sparse) matrix, where $|n\rangle$'s are the eigenstates of $H$.
      The results are shown in Fig.~\ref{fig1}, where we use the median of the scaled decay rates as the ``typical'' value, with error bars estimated based on the 68 percent bootstrap confidence interval.
        For weak bath coupling $\gamma=0.001$, the perturbation theory produces almost identical results to the ED. For strong bath coupling, the agreement obviously degrades but we can still see the intersections manifesting consistently in both methods.
        As such, we use the first-order perturbation in $\gamma$ for the discussion of the decay rates below, calculated up to $L=14$, displayed in the scale of $4^{-L}\gamma$ (since $1/\tau\propto\gamma$). 
 For the quasiperiodic model, data with $L\leq9$ are calculated from 6000 random choices of the initial phase $\phi$. For larger $L$, the number of random choices is adaptive and varies around 100 to 1000.
	
      Another issue arising as we calculate the decay rate with larger $L$ is the floating-point error.
      When the decay rate become so small, comparable to the machine epsilon of the double-precision floating-point arithmetic, the value will become very inaccurate.
      To mitigate this issue, we follow Refs.~\cite{Sels2021,Morningstar2022} to use the 80th percentile as the ``typical'' decay rate, instead of the median, since a larger percentile is more away from the machine epsilon.
It has been discussed in Ref.~\cite{Morningstar2022} that choosing different percentiles does not cause qualitative differences.
Even so, some data points of the ``typical'' decay rate are still too close to the machine epsilon. 
In order to maintain the validity of our data, we drop the data points with $\gamma^{-1}P_{80}(1/\tau)<100\epsilon$, where $\epsilon=2.2\times10^{-16}$ is the machine epsilon. 
The choice of $100\epsilon$ is justified by the observation that the statistical error of our data is at the order of $10\%$, so that we can neglect the arithmetic error above $100\epsilon$, which is roughly $\lesssim1\%$.
The error bars in the figures below are estimated based on the 68 percent bootstrap confidence interval.

	\begin{figure}
	\centering
	\includegraphics[width=0.75\columnwidth]{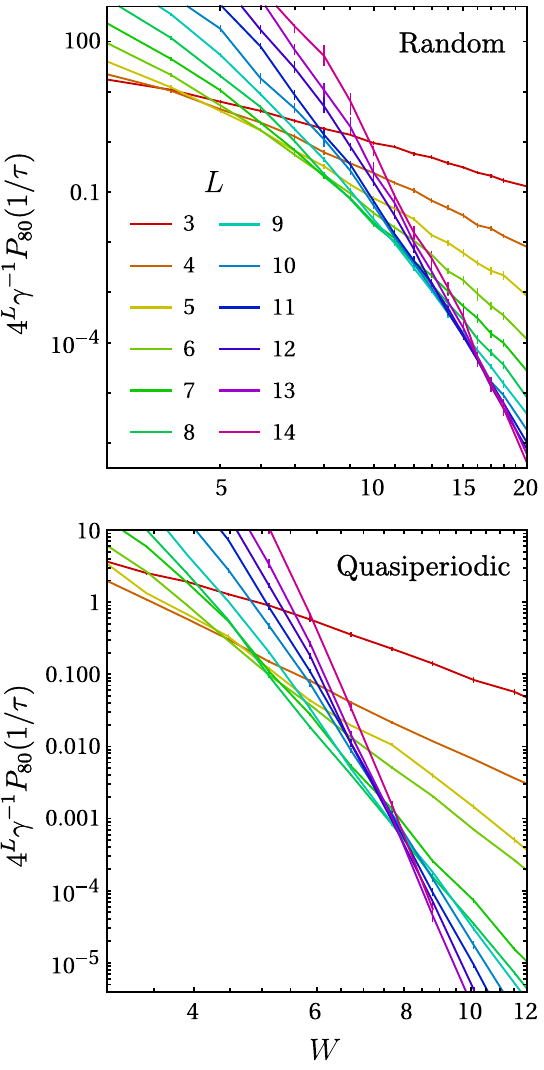}
        \caption{Scaled decay rates versus the disorder strength $W$ at different system length $L$ for the random and the quasiperiodic spin chain. Data for the random model with $L\geq5,W\geq14$ are taken from Ref.~\cite{Morningstar2022}. }
	\label{fig:ratevsw}
   \end{figure}	
	
   \subsection{Results}
\begin{figure}
	\centering
	\includegraphics[width=0.45\textwidth]{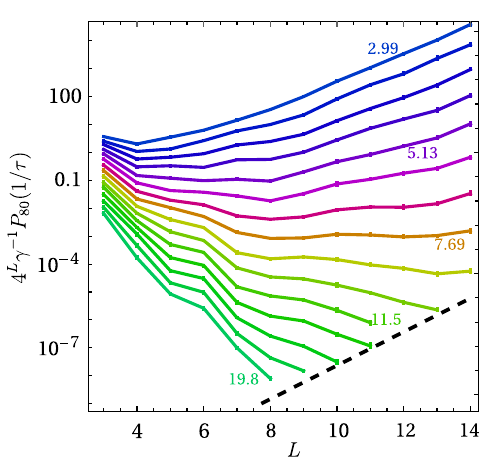}
        \caption{Scaled decay rates versus the length $L$ of the quasiperiodic spin chain for a given disorder strength $W$. The value of $W$ increases logarithmically from the top to the bottom lines, and we mark several values of $W$ for reference.  Dashed line indicates $\gamma^{-1}P_{80}(1/\tau)=100\epsilon$ with $\epsilon$ being the machine epsilon of the double-precision floating-point numbers. We drop data points below this line. }
	\label{fig:rateVsL}
   \end{figure}

   In Fig.~\ref{fig:ratevsw}, we show the scaled ``typical'' decay rate in the random and the quasiperiodic spin chains.
For the random model, the curves drift smoothly as $L$ grows larger, without any clear common intersections of the curves (within the error bars).
On the other hand, the result of the quasiperiodic model is qualitatively different, with a common intersection at $W_c\sim 7-8$ emerging for $L\geq8$.
As a result, for $W < W_c$, the decay rate varies with $L$ faster than $4^{-L}$, indicating an avalanche instability.
The quite small sizes of our simulation naturally raise a question of whether these intersections really converge to a finite $W_c$ or they drift to infinity, ruling out any true MBL.
To determine how the avalanche landmark evolves in large systems, we plot in Fig.~\ref{fig:rateVsL} the decay rate with respect to $L$ for the quasiperiodic model. Here, a decreasing (increasing) function indicates MBL (ETH) while the minimum marks the finite-size avalanche criticality. For $W>8$, within numerical fluctuations, the scaled decay rates decrease monotonically for all values of $L$, implying that $W\sim 7-8$ is an overestimation for avalanche instability in the thermodynamic limit; while for $W=7.69$, a broad minimum emerges starting at $L=10$, indicating that the avalanche landmark most likely saturates for $L\geq 10$.

\begin{figure}
	\centering
	\includegraphics[width=0.45\textwidth]{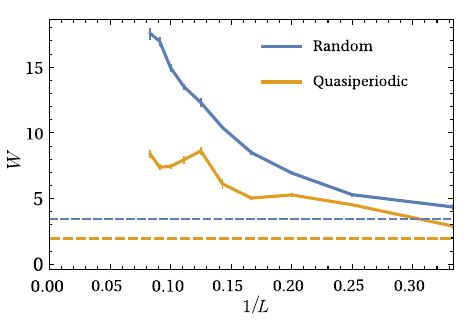}
        \caption{The value of $W$ at which the spin chain of length $L$ and $L+2$ have the same $4^L/\tau$. Data of the random model with $L\geq 5, W\geq 14$ are taken from Figure 5 of Ref.\ \cite{Morningstar2022}. The dashed lines denote the typical finite-size ($L=14-16$) MBL transition of the respective model (color-wise) from the level statistics (Fig.~\ref{levelstatistics}).
	}
	\label{fig:intersections}
    \end{figure}

In Fig.~\ref{fig:intersections}, we plot the intersecting $W$ with the system length $L$ for the random and the quasiperiodic spin chains.
Data points are calculated by fitting each $L$ and $L+2$ curves near their intersection with quadratic functions in the $\log\left(4^LP_{80}(1/\tau)\right)$-$\log W$ plane, with the intersection point as the shared parameter. For the random model, the intersection drifts continuously with $L$ but appears to slow down near $L=12, W=17$. From this fact, Ref.~\cite{Morningstar2022} postulates a possible thermodynamic avalanche landmark. For our quasi-periodic model, this indicator is much sharper as one can see that the convergence of the intersections starts at smaller $L$ and lower value of disorder strength $W$ than the random spin chain. We note that the slight fluctuation visible in Fig.~\ref{fig:intersections} is a finite size effect caused by imposing an irrational period on a finite lattice. See Appendix~\ref{app:open} for more discussion.
	
    The $L=14-16$ values of the finite size MBL transition (see Fig.~\ref{levelstatistics}) are shown as the dashed lines in Fig.~\ref{fig:intersections}, to compare with the avalanche landmark.
Note that the avalanche landmark is much closer to the finite-size MBL transition obtained through the level statistics data for the quasiperiodic spin chain. This is in contrast to the random case where the avalanche instability can happen deep in the finite-size MBL phase \cite{Morningstar2022}, thus raising some questions about the thermodynamic MBL. 
Our bath coupling results, on the other hand, suggest that the extreme case of avalanche (induced by an ideal bath) in the quasiperiodic model is
less severe than in the random model, possibly up to arbitrarily large system sizes.

    \section{Real-space renormalization}\label{sec:rg}

	\begin{figure}
	\includegraphics[width=0.48\textwidth]{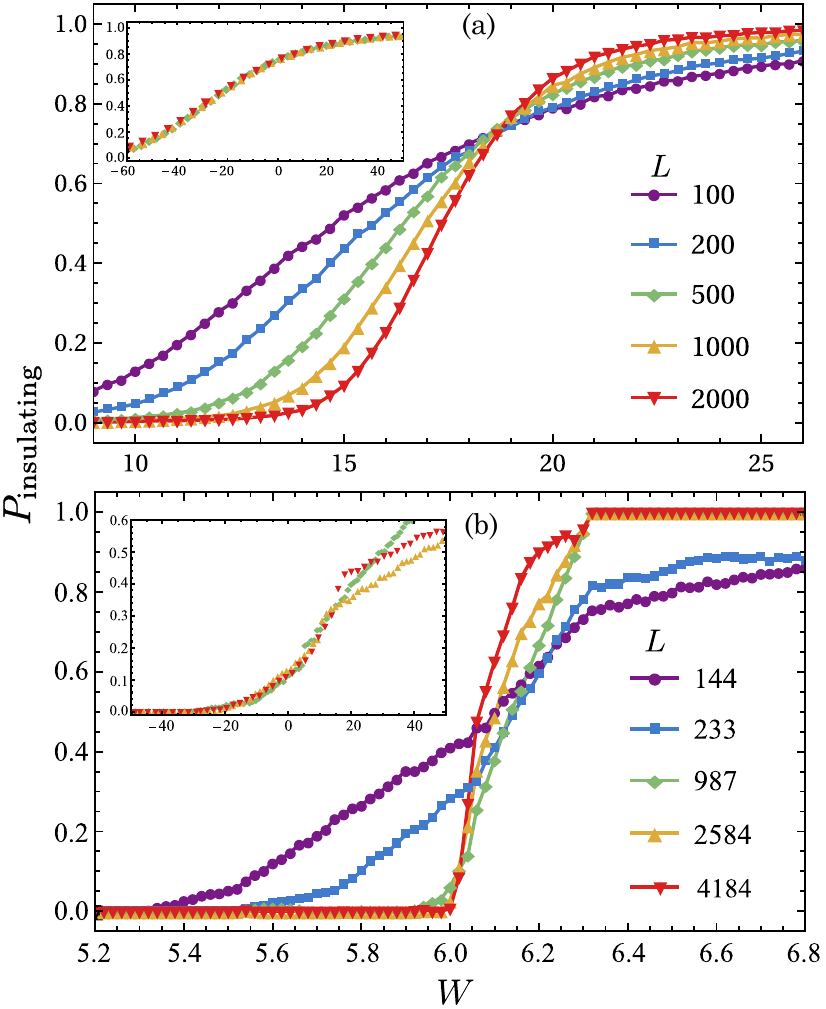}
	\caption{Probability of renormalized system being insulating for the random (a) and quasi-periodic (b) onsite potential. The inset shows the respective collapsed data as a function of $(W-W_c)L^{1/\nu}$ with $W_c=18.9$, $\nu=2.8$ (a) and $W_c=6.0$, $\nu=1.2$ (b).\label{figRG}}
\end{figure}
	\begin{figure}
	\includegraphics[width=0.48\textwidth]{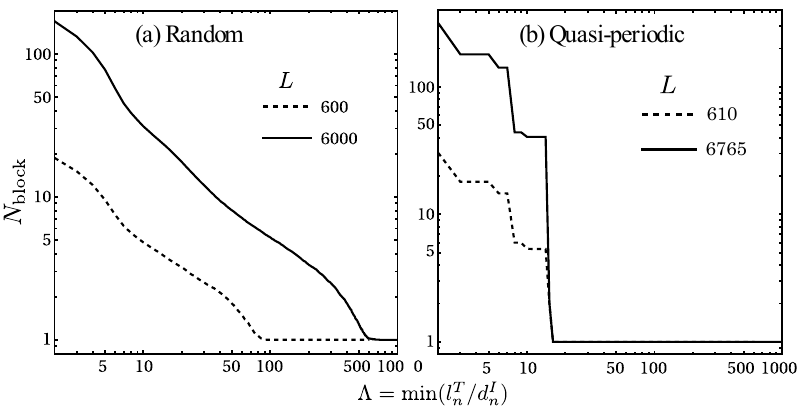}
	\caption{Number of present clusters after decimating all the blocks shorter than the cutoff length $\Lambda$. The disorder strength $W=17$ for the random model (a) and $W=5.5$ for the quasi-periodic model (b).\label{figRG2}}
    \end{figure}
   
    \begin{figure*}
     \centering
     \includegraphics[width=0.9\textwidth]{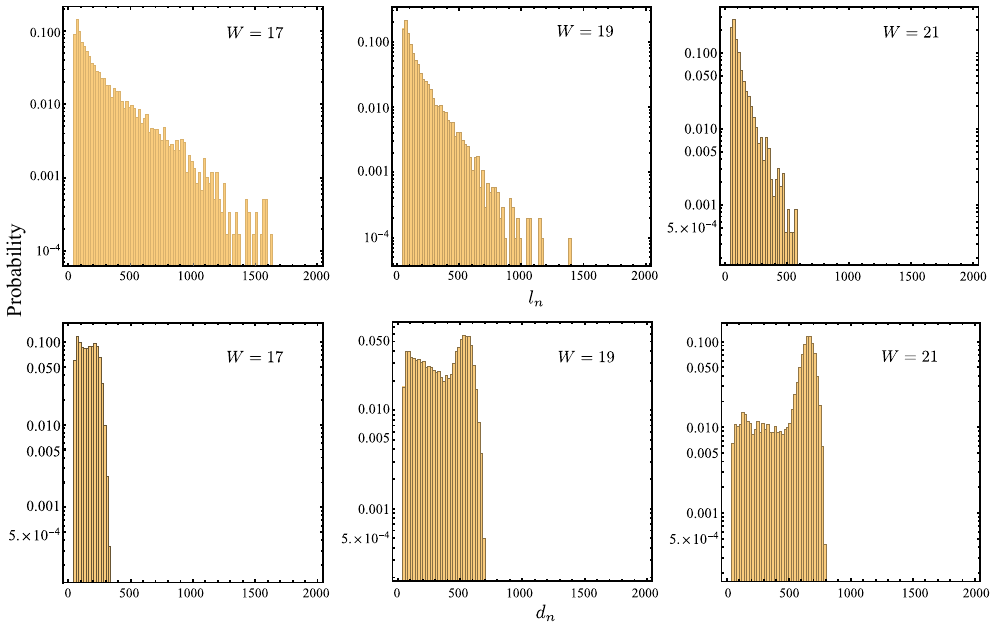}
     \caption{Probability density for thermal block physical length (top panel) and insulating block primary length (bottom panel). The system size is $L=6000$ and blocks are decimated until $l_n/d_n\geq 50$. \label{fig3}}
    \end{figure*}

	One might think that the difference between the two models is simply quantitative and due to a specific choice of the quasi-periodic potential instead of a more general one, i.e. $h(x)=h\left(x+\varphi^{-1}\right)$. In the following discussion, we establish a stronger scenario and demonstrate that they belong to different universality classes. As such, MBL in quasi-periodic systems is more stable (in the thermodynamic limit) than in the random systems. 
	
	To show this, we use the strong randomness real-space renormalization \cite{Zhang2016,Goremykina2019,Morningstar2019} where the initial state of the 1D spin chain is composed of alternating thermal (T) and insulating (I) blocks.
        Here, the T segments are only characterized by their physical length $L^T$ while the I blocks, beside the physical length $L^I$, have extra index $d^I=L^I/\xi-L^I$, called the primary length, that indicates how close the block is to avalanche instability.
        $\xi$ is normalized so that $\xi<1$ is associated with localization, so $d^I$ is always positive. 
        The following rules of block decimation apply.
        If $L^T_i < d^I_{i-1}, d^I_{i+1}$, the three-block segment merges into a new insulating block with $L^{'I}_{i} = L^I_{i-1}+L^T_i+L^I_{i+1}$ and $d^{'I}_i = d^I_{i-1} -L^T_i+ d^I_{i+1}$.
        Otherwise, if the smallest block (in primary length) is insulating, i.e.\ $d^I_i<L^T_{i-1},L^T_{i+1}$, the composite block becomes thermal with $L^{'T}_i=L^T_{i-1}+L^I_i+L^T_{i+1}$.

        The details for the RG flow in random and quasi-periodic models are described in Refs.~\cite{Morningstar2019,Morningstar2020,Agrawal2020} but the initial block configuration is too simple for our microscopic models. 
        On the other hand, there are RG schemes for generic spin chain models, but the procedure is much more intricate \cite{Imbrie2016,Imbrie2016b,Thiery2018}.
For our purpose, it is important to treat both random and quasiperiodic models on the same footing, so we aim at performing the strong randomness renormalization starting from the microscopic Hamiltonian in Eq.~\eqref{eq:quasiperiodic_hamiltonian},
where the details are described below.
        Note that although we do not prove explicitly that this infinite-randomness RG procedure is asymptotically correct for our quasiperiodic spin chain, we will show below that it gives the same universality class as the quasiperiodic model of Ref.~\cite{Agrawal2020} which can be proved analytically to be asymptotically correct. This justifies our coarse-graining approach.

\subsection{Initialization of RG flow}
        
        Our approach of initializing $\xi$ is inspired by the connection between the off-diagonal coupling and link resonance.
        In the product state basis $\ket{\eta}=\prod_{j=1}^{L} \otimes \ket{\eta_j}$ where $\eta_j\in \mathbb{C}^d$ is one of the basis of the local $d-$ dimensional Hilbert space at site $j$. Assuming only nearest coupling, the Hamiltonian can be written as
    \begin{equation}
    	H=\sum_j J_{j,j+i} + U_{j,j+1} + V_j
    \end{equation}
    where $J$ and $U$ have support on pairs of nearest sites, and the previously defined product state basis diagonalizes $U$ and $V$. Since only $J$ has off-diagonal matrix elements, it introduces a coupling between product states $\ket{\eta}, \ket{\eta'}$ defined by the ratio
    \begin{equation}
    	G_{j,j+1}(\eta,\eta') =\left | \frac{\bra{\eta}V_{j,j+1}\ket{\eta'} }{E_\eta-E_\eta'} \right|
    \end{equation}  
    with $E_\eta=\bra{\eta}U+V\ket{\eta}$. For the spin chain model discussed in the main text, we can identify
    \begin{equation}
      \begin{gathered}
    	J_{j,j+1}\equiv \frac{1}{2}(\sigma^\dagger_j\sigma^-_{j+1} + \sigma^\dagger_{j+1}\sigma^-_j),\\
        U_{j,j+1}\equiv \frac{1}{4} Z_{j}Z_{j+1}, \quad V_j\equiv \frac{1}{2} h_jZ_j.
      \end{gathered}
    \end{equation}
    At one particular link, we can have four off-diagonal coupling $\ket{\uparrow \textcolor{red}{\uparrow \downarrow}\uparrow}\rightleftarrows \ket{\uparrow \textcolor{red}{\downarrow \uparrow}\uparrow}$, $\ket{\uparrow \textcolor{red}{\uparrow \downarrow}\downarrow}\rightleftarrows \ket{\uparrow \textcolor{red}{\downarrow \uparrow}\downarrow}$, $\ket{\downarrow \textcolor{red}{\uparrow \downarrow}\uparrow}\rightleftarrows \ket{\downarrow \textcolor{red}{\downarrow \uparrow}\uparrow}$, and $\ket{\downarrow \textcolor{red}{\uparrow \downarrow}\downarrow}\rightleftarrows \ket{\downarrow \textcolor{red}{\downarrow \uparrow}\downarrow}$ with the red spin indicating the link of interest and other spins beyond the two nearest ones are factored out. Following Ref.~\cite{Thiery2017}, we take the geometric means of these dimensionless couplings, obtaining
    \begin{equation}
    	\bar{G}_{j,j+1} =  \left( \frac{1}{2\Delta h_j } \frac{1}{2\Delta h_j } \frac{1}{2\left|\Delta h_j+1 \right|}
    	\frac{1}{2\left|\Delta h_j-1\right|}\right)^{1/4}
    \end{equation}
    Instead of comparing $\bar{G}$ with unity to define the resonant or perturbative link, we benchmark it against $1/2$ (thus a stronger condition).
    This $1/2-$threshold tunes $P_\text{insulating}$ at the critical point not too close to either 0 or 1, allowing better visualization of data collapse.
    
    Finally, the $\xi$ parameter for the link is given by
    \begin{equation}\label{eqxi}
    	\xi_{j,j+1} = \left(2\bar{G}_{j,j+1}\right)^{\alpha} = \left( \frac{1}{\Delta h_j^2 } \frac{1}{\left|\Delta h_j+1 \right|}
    	\frac{1}{\left|\Delta h_j-1\right|}\right)^{\alpha/4},
    \end{equation}
    with $\xi < (\geq)1$ indicating the insulating (thermal) link.
    Here we introduce a parameter $\alpha$ to adjust the critical disorder strength so that it agrees with the open system simulation in Sec.~\ref{sec:open}, which we found to be $\alpha=0.1$ (note that we use a single parameter to fit both the random and the quasiperiodic models).
The free parameter $\alpha$ does not change the universality class and qualitative features of the RG flow but only affects the exact value of the critical point (see Appendix \ref{app:rg}).
     For insulating links, we can define the primary length $d_{j,j+1}=1/\xi_{j,j+1}-1$. After characterizing all links, we group similar-type links into a block whose physical and primary length is given by the sum of component links.

        Through Eq.~\eqref{eqxi}, we can use the exact microscopic model as the initial configuration for RG.
        The purpose of this analysis is obviously not predicting the critical disorder strength, but comparing the two models and highlighting the different natures of their seemingly similar ETH-MBL transitions.
        
\subsection{Results}\label{sec:rgdetails}

  \begin{figure*}
	\centering
	\includegraphics[width=0.9\textwidth]{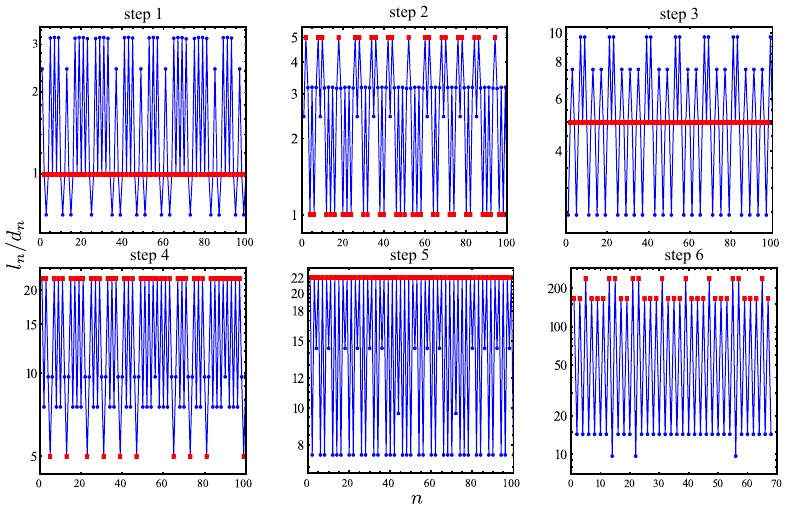}
	\caption{Physical/Primary length of thermal (red) and insulating (blue) blocks with respect to the block order index. We simulate a system with size $L=10946$, $\phi=0$, and $W=5.5$. \label{fig5}}
   \end{figure*}    
	
In Fig.~\ref{figRG}, we show the probability that the entire system becomes insulating after coarse-graining, where we sample different random configurations for the random model, and sample different initial phases for the quasiperiodic model.
Note that we choose the system size to be Fibonacci numbers in the quasi-periodic case for better compatibility with the periodic boundary condition. 
For a generic system size, the situation does not change significantly but the data collapse is slightly worse. 
The critical disorder strengths in Fig.~\ref{figRG} for the random and quasi-periodic models, i.e. $W_c=19$ and $6$ respectively,  agree with the avalanche critical values obtained from studying the bath-coupled systems. Remarkably, one can contrast the transition of these two MBL models. For the random potential, the severe finite-size effect is visible even for the system as large as 2000 sites, making the ETH-MBL transition manifestingly a glassy crossover regime over a wide range of $W$. This means that MBL in random systems would show severe finite-size effects with slow relaxation and the putative critical disorder drifting upward with increasing system size. By contrast, the quasi-periodic model exhibits a much sharper transition [comparing the $W$ scales of $1.6$ of Fig.~\ref{figRG}(b) with the scale of $17$ of Fig.~\ref{figRG}(a)] as well as a faster convergence to the thermodynamic limit. Additionally, we can collapse the finite-size data into a universal function of the form $f[(W-W_c)L^{1/\nu}]$, yielding the scaling exponent $\nu=2.8$ and $1.2$ for the random and quasi-periodic models, respectively. This suggests that the two models in our study belong to different universality classes. In fact, for the former case, the exponent agrees with the value $2.6$ in Ref.~\cite{Morningstar2019} where the initial block configuration is sampled randomly from a predetermined distribution. For the latter, our numerically obtained $\nu$ is consistent with the exact value of unity derived from the enforced quasi-periodic configuration $l_n\sim \cos(2\pi n \varphi)$ \cite{Agrawal2020}. It is remarkable that even though Eq.~\eqref{eqxi} is defined somewhat arbitrarily and similar links are first coalesced into a cluster, our two RGs still preserve the characteristics of their respective universality classes. More importantly, we are able to introduce the physical disorder strength into the renormalization, validating our open system results that the avalanche critical disorder strength for the quasi-periodic model saturates much sooner and faster with respect to $L$ and at lower value of $W$. Thus, finite-size MBL is much sharper and thermodynamic MBL is much more likely in the quasiperiodic model than in the random model.

   More visible distinctions between the two models are revealed in the details of the RG flow. By decimating blocks, the smallest cluster length progressively increases until it reaches the system size, so we treat this quantity as the flow parameter. In Fig.~\ref{figRG2}, we show the average number of blocks left as we increase the cutoff length.  For the random model, there exists a smooth tail whose extension increases with the system size. 
   On the other hand, the cluster decimation in the quasi-periodic model happens abruptly and saturates at a length ($\sim 20$) uncorrelated with and much shorter than the system size.
    We demonstrate below that this qualitative difference is due to the different cluster length distribution: the random model has a broad distribution, where exceptionally large cluster is statistically possible for large enough system size; while in the quasiperiodic model, clusters have their lengths concentrated into narrow bands and thus are decimated almost simultaneously.

    To demonstrate the broad distribution for the random model, we show in Fig.~\ref{fig3} the distribution of thermal and insulating blocks in the random model as we increase the disorder strength passing the critical point. Both types of blocks exhibit continuous broad distributions, with large thermal regions being exponentially rare. As $W$ increases, the distribution for insulating blocks extends and surpasses that of the thermal blocks for $W > W_c$.

Now we demonstrate the block parameter sharpening for the quasiperiodic mode. First, we note that
if we instead approximate the irrational $\varphi$ by $F_{k+1}/F_k$ with $F_k$ being the $k$th Fibonacci number, the link parameters repeat itself after $F_k$. After grouping similar-type links into blocks, the block parameter is also periodic after every $\kappa(F_k)$ block with $\kappa(F_k)$ being an integer less than $F_k$ because of the link merging. We expect that as $k\to \infty$, $\kappa(F_k)\to \infty$, so the initial block configuration does preserve quasi-periodicity even though we do not know its explicit position dependence. This places our model into the same universality as Ref.~\cite{Agrawal2020}. Therefore, we expect the block parameters to sharpen along the RG flow. In Fig.~\ref{fig5}, we denote the initial block parameters as step 1 (the x-axis is the block index). We can see there exist insulating blocks both longer and shorter than the thermal blocks. However, these lengths group into approximate bands, in which the lowest band only contains local minimum (block shorter than its two immediate neighbors). Therefore, if we increase the cutoff length beyond this band, the whole band is eliminated simultaneously. As we apply this procedure iteratively, at step 6, all the thermal and insulating blocks already concentrate into two respective bands well separated from each other. It is noticeable that throughout the RG flows, block lengths only assume a few values much shorter than the system size, instead of forming a broad distribution like the random case. 
This is a direct manifestation of the rather weak finite size effects in the quasiperiodic model, particularly compared with the random model where finite size effects are severe.
    
   To conclude, we contrast the MBL stability in the two models. For the random model, large thermal regions can emerge and weaken the localization. Due to the statistical nature of these rare regions, as the system size grows, larger thermal regions also occur making the glassy behavior persist to a very large size. For the quasi-periodic model, instead of growing to exceptionally large size (albeit low probability), the sizes of all thermal spots mostly assume only a few values, so they either all become inactive in the MBL side or all expand in the thermal size. As a result, the MBL is more stable and finite-size effect is less severe. Our findings are consistent with the recent state-of-the-arts numerical experiments finding that the power-law decaying imbalance persists in the interacting Anderson model for large system sizes and strong disorders \cite{Sierant2022}. This behavior does not occur in the quasi-periodic model. 
   
\section{Conclusion}\label{sec:conclusion}
We have studied the avalanche instability in an interacting quasiperiodic chain by coupling to a thermal bath, finding an unexpected stability of MBL for disorder strength $W>8$, implying that thermodynamic MBL is likely more stable in quasiperiodic systems for moderate disorder, in contrast to the randomly disordered chains where thermodynamic MBL does not exist up to a disorder strength of 18.
    In addition, the real space RG approach shows qualitative differences between the distribution of rare thermal regions in the two models.
    The rarity of large thermal seeds in the quasiperiodic spin chain implies that the worst-case scenario in our open system simulation is not even likely to occur in a long spin chain, strengthening the conclusion that quasiperiodic potentials are more stable.
    Our work suggests that MBL experiments in many-atom ($>$100) 1D systems may benefit from quasiperiodic potential.
    An interesting future direction is to extend the real space RG approach to an analytically more rigorous treatment, in particular, a more direct way to begin the RG from the microscopic model.

\section*{Acknowledgement} This work is supported by the Laboratory for Physical Sciences. Y.-T.~T. and D.~V. contributed equally to this work. 
   
\bibliographystyle{apsrev4-2}
   \bibliography{references}

\onecolumngrid
   \appendix

   \section{Additional open-system simulation data}\label{app:open}

\begin{figure}[h]
    	\centering
    	\includegraphics[width=0.6\textwidth]{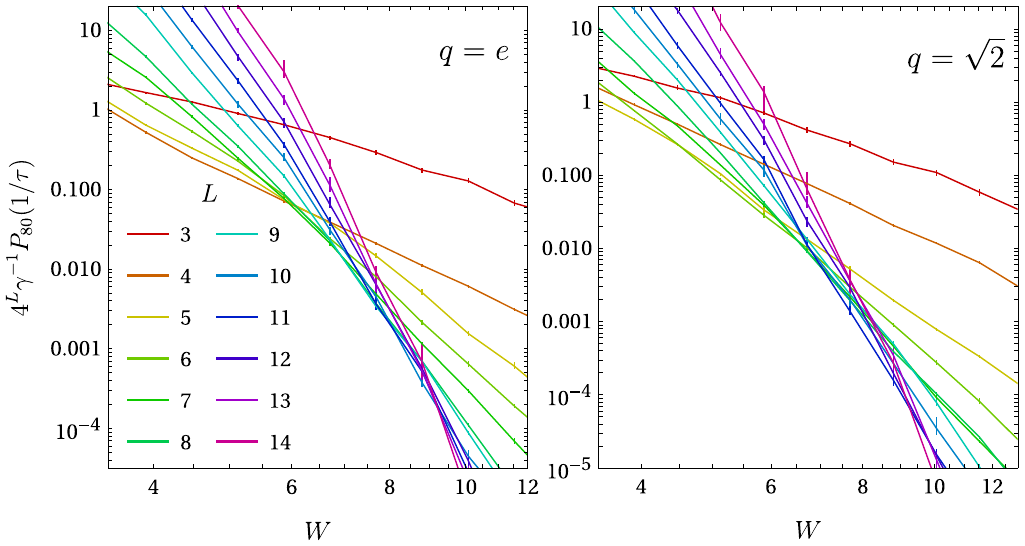}
        \caption{Scaled decay rate with respect to $W$ for the random model and the quasiperiodic models with quasiperiod $q=e$ and $\sqrt{2}$. \label{fig2} }
    \end{figure}

\begin{figure}[h]
    	\centering
    	\includegraphics[width=0.7\textwidth]{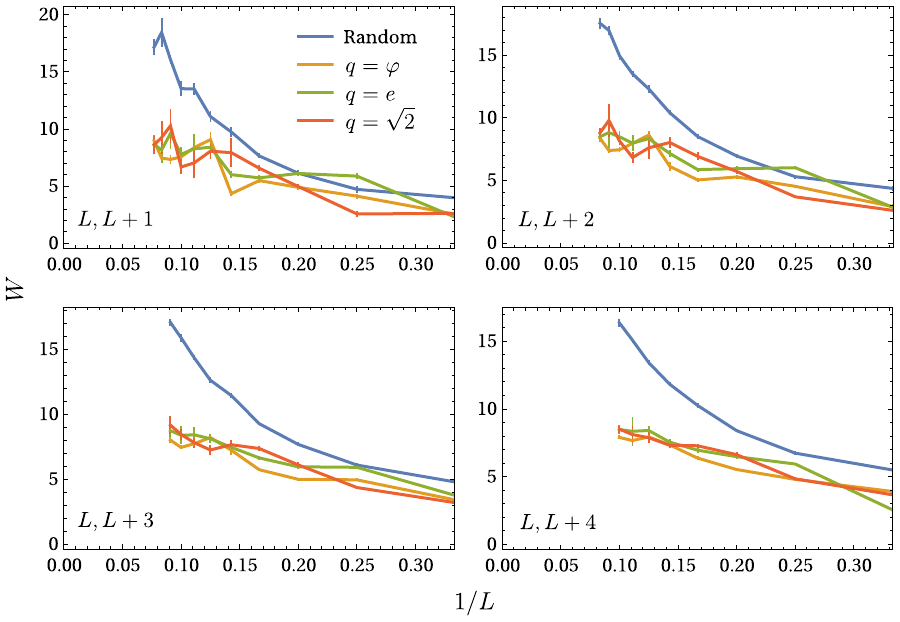}
        \caption{ Intersecting disorder strength between $L$ and $L+n$ for the curves in Figs.~\ref{fig:ratevsw} and \ref{fig2}. \label{fig2-1}}
    \end{figure}

        In Fig.~\ref{fig2}, we show the scaled decay rate with respect to $W$ for the quasiperiodic models with quasiperiod $q=e$ and $\sqrt{2}$ ($h_j=W\cos(2\pi q j+\phi)$), as a complementary to the $q=\varphi$ in Fig.~\ref{fig:ratevsw}.
      For the random model (Fig.~\ref{fig:ratevsw}), all the lines are regular while for the quasiperiodic models, there exist sudden jumps at small $L$ that depend on $q$. At larger $L$, the data become more regular and uniform across different $q$.
        The intersections are shown in Fig.~\ref{fig2-1}. We can see that the intersections for quasiperiodic models are around $W\sim 7-8$, independent of the quasiperiod $q$. On the other hand, the slight fluctuating pattern around this value clearly varies with $q$ and is a finite-size effect due to imposing an irrational quasiperiod on a finite lattice causing some correlated non-uniformity. We expect this effect to diminish at larger sizes.  On the other hand, the intersecting points of the random model visibly drift with increasing system size. 
    

\begin{figure}[h]
    	\centering
    	\includegraphics[width=0.7\textwidth]{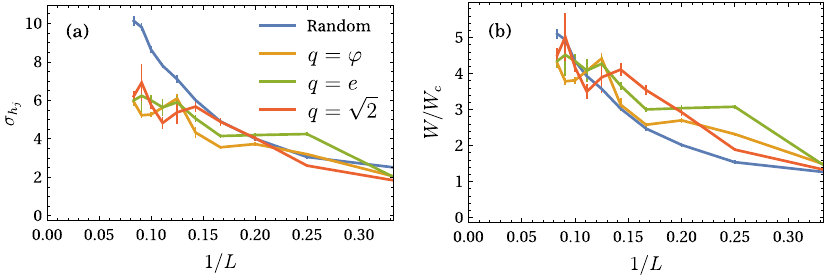}
        \caption{ (a) Intersecting disorder strength, measured in the standard deviation of a single $h_i$, between $L$ and $L+2$ for the curves in Figs.~\ref{fig:ratevsw} and \ref{fig2}.
          (b) Intersecting disorder strength divided by the finite-size ($L\sim 16$) critical disorder strength (Fig.~\ref{levelstatistics}).
          \label{fig:scale}
        }
    \end{figure}

    One may be concerned about whether comparing the $W$ of the random and the quasiperiodic spin chains is meaningful. In particular, the distribution functions of a single $h_j$ are different, being the uniform distribution for the former, and the arcsine distribution for the latter.
    Here we stress that there is no unique quantitatively meaningful way to compare the disorder strengths of the random and quasiperiodic systems, and the parameter $W$ used in this paper is not for a precise quantitative comparison between the two models. Although we make some rough comparisons between their numerical values, a more important point is to show the qualitative difference between the two models.
    Another possible parameter to compare the two models is the standard deviation of a single $h_j$, which equals $W/\sqrt{3}$ for the random model and $W/\sqrt{2}$ for the quasiperiodic model.
    The result is shown in Fig.~\ref{fig:scale}(a). Note that this does not change the main point that the quasiperiodic system is more likely to exhibit thermodynamic MBL.
    In addition to comparing the values of the avalanche landmarks themselves, one may also be interested in the ratio of the avalanche landmarks to the critical strengths of the corresponding finite-size systems, which is independent of the choice of the parameter for disorder strengths. The result is shown in Fig.~\ref{fig:scale}(b).

  \section{Effect of the tuning parameter in the real-space renormalization}\label{app:rg}
\begin{figure}
    	\centering
    	\includegraphics[scale=0.56]{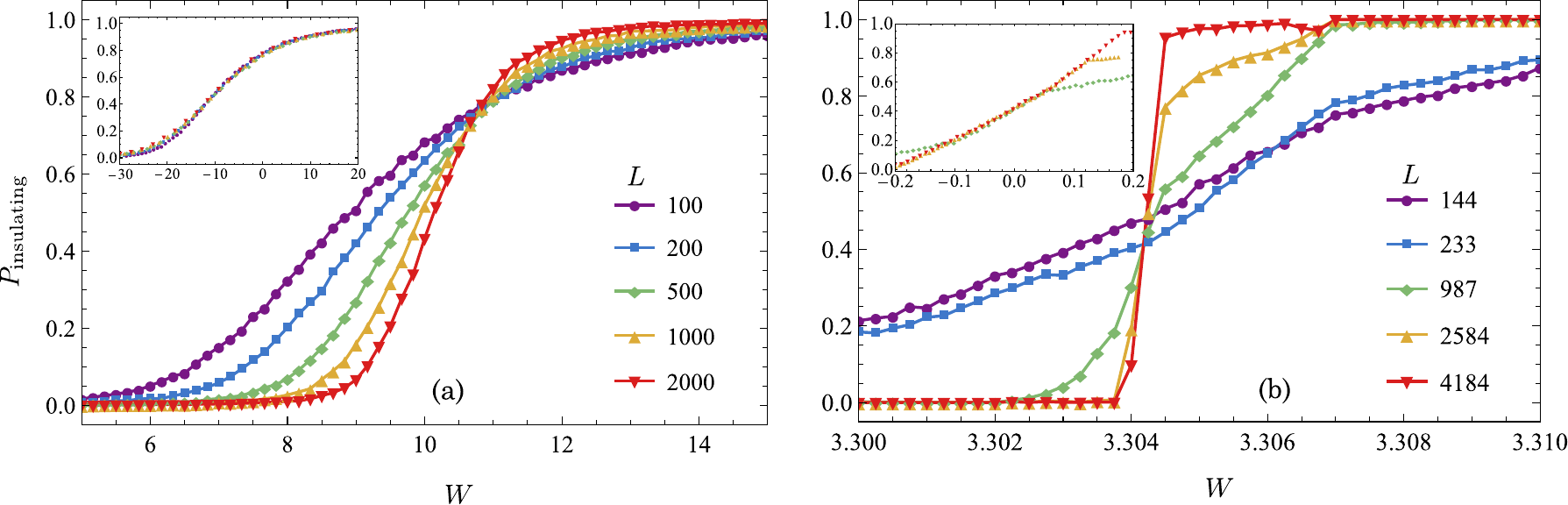}
    	\caption{Finite-size scaling of $P_{\text{insulating}}$ with $\alpha=0.18$ in Eq.~\eqref{eqxi} for the random (a) and quasiperiodic (b) models. The inset shows the collapsed data with respect to $(W-W_c)L^{1/\nu}$ where $(W_c,\nu)=(10.8,2.9)$ for the random model and $(W_c,\nu)=(3.3,1.2)$. \label{fig:alpha18}}
    \end{figure}

    We note that the tuning parameter $\alpha$ only affects the value of the critical $W$ without changing the universality class or qualitative difference between the random and the quasiperiodic models. In Fig.~\ref{fig:alpha18}, we show the RG results with the tuning parameter $\alpha=0.18$ instead of $\alpha=0.1$ used in the main text. The finite-size scaling exponents are unaffected in both models, and the insulating-conducting transition for the random model still happens at a much higher value of disorder strength (10.8 versus 3.3 compared to 18.9 versus 6.0 for $\alpha=0.1$) and over a much wider range of $W$ than the quasiperiodic model.
    
\end{document}